\documentstyle[psfig,12pt]{article}
\begin{document}
\title{Self-consistency in the Projected Shell Model} 

\author{
V. Vel\'azquez$^1$, J. G. Hirsch$^2$\thanks{On sabatical leave from
Departamento de F\'{\i}sica, Centro de Investigaci\'on y
Estudios Avanzadas del IPN.}
 and Y. Sun$^3$\\
{\small\it $^1$ Departamento de F\'{\i}sica, Centro de Investigaci\'on y
Estudios Avanzados del IPN,} \\
{\small\it A.P. 14-740, 07000 M\'exico D.F., M\'exico    }  \\
{\small\it $^2$ Instituto de Ciencias Nucleares, UNAM,} \\
{\small\it Circuito Exterior C.U.,A.P. 70-543, 04510 M\'exico D.F.,
M\'exico    }  \\
{\small\it $^3$Department of Physics and Astronomy, University of Tennessee,} \\
{\small\it Knoxville, Tennessee  37996, USA} \\
} 
\maketitle

\begin{abstract}
The Projected Shell Model is a shell model theory built up over a
deformed BCS mean field. Ground state and excited bands in even-even
nuclei are obtained through diagonalization of a pairing plus quadrupole
Hamiltonian in an angular momentum projected 0-, 2-, and 4-quasiparticle basis.
The residual quadrupole-quadrupole interaction strength is fixed
self-consistently with the deformed mean field and the pairing constants are
the same used in constructing the quasiparticle basis.
Taking $^{160}Dy$ as an example, we calculate low-lying 
states and compare them with experimental data. We 
exhibit the effect of changing the residual interaction
strengths on the spectra. It is clearly seen that there are many
$J^\pi = 0^+, 1^+, 4^+$ bandheads whose energies can only be reproduced  
using the self-consistent strengths. It is thus concluded that the 
Projected Shell Model is a 
model essentially with no free parameters.
\end{abstract}



\section{Introduction}

It is known that 
the shell model is the most fundamental way of
describing many-nucleon systems fully quantum
mechanically. However, using the shell model to study
deformed heavy nuclei is a desirable but very difficult
task because of large dimensionality and its related
problems. 
Therefore, for studying the heavy nuclear systems, one 
has had to rely mainly on algebraic or mean-field type models. 

The Projected Shell Model (PSM) \cite{Har95}  
provides one possible solution for
a shell model treatment in heavy systems. 
In this approach, one first truncates
the configuration space with guidance from the deformed
mean field by selecting only the BCS vacuum plus a few
quasiparticle configurations in the Nilsson orbitals
around the Fermi surface, performs angular momentum
projection (and particle number projection if neccessary)
to obtain a set of laboratory-frame basis
states, and finally diagonalizes a shell-model
Hamiltonian in this space. Since the deformed mean field
+ BCS vacuum already incorporates strong particle-hole
and particle-particle correlations, this truncation
should be appropriate for the low-lying states dominated
by quadrupole and pairing collectivity. 

Indeed, this
approach has been very successful for ground band
properties and near-yrast quasiparticle
excitations in high-spin physics for both
normally deformed \cite{Har95} and superdeformed states \cite{Sun97}
in heavy nuclei.
It can describe the spectra and electromagnetic
transitions quantitatively. 
Using microscopic degrees of freedom, it is able to explain many
high spin phenomena. 
Very recently, attempts of using this model to 
study double-beta decay \cite{Hirsch97}
and the physics away from the beta stability region 
\cite{Zhang98} have been made. 

To compare with a conventional shell model space, the PSM is a 
severely truncated theory 
with space dimension typically being $10^2$. 
However, the truncation in the PSM is done in a very appropriate way.
We know that in treating a deformed system, a simple configuration in
a conventional shell model does not correspond to any simple
mode of excitation. This means that a huge configuration space
would be neccessary just to represent even the lowest eigenstate 
(the ground state) of the Hamiltonian. In the PSM, on the other hand,
the ground state band is obtaind by projecting on the vacuum state.
Thus a single projected configuration in the PSM 
can be equivalent to
a linear combination within a large configuration space in a 
conventional shell model.  

Of course, one could ask a question of how effective interactions 
in this severely truncated theory affect the final results. 
In the present article, it will be shown that, while in practice very few
multi-quasiparticle states are included in the diagonalization, any
departure from the interactions in the Hamiltonian with 
self-consistent values destroys the agreement between theory and
experiment. After reviewing the model and the self-consistent
determination of the quadrupole-quadrupole interaction strength 
we calculate low-lying bands in $^{160}Dy$ and exhibit the effect of
changing the residual interaction strengths on the spectra. It is
demonstrated that most bandheads energies can
only be reproduced using the self-consistent strengths. We thus 
conclud
that the PSM has essentially no free parameters.

\section{The Model}

The PSM is built over a deformed mean field, which incorporates
pairing effects through a Bogolyubov transformation to
quasiparticle states. Both the Nilsson deformation parameter $\epsilon$
and the pairing constants $G_p$ and $G_n$ are taken from systematics
\cite{Har95}.
The projected BCS vacuum gives the unperturbed ground
state band in even-even nuclei, while unperturbed excited bands are
obtained by projecting the multi-quasiparticle states. 
The multi-quasiparticle
states are those with two proton and two
neutron quasiparticle excitations (for even-even nuclei), with
proton-neutron quasiparticle pairs (for odd-odd nuclei), or with one and
three quasiparticle states (for odd nuclei).
Angular momentum projection is exactly performed, providing the building
blocks of the model. 

A shell model Hamiltonian is diagonalized in this basis. It contains the
single particle energies, monopole pairing between like particles,
quadrupole-quadrupole and quadrupole-pairing interactions. 
The residual quadrupole-quadrupole interaction is fixed self-consistently
with the deformed mean field and the pairing constants are the same used 
in building up the quasiparticle basis. The only remaining parameter is
the one associated with the quadrupole pairing interaction, and it is set
as 1/5 of the monopole-pairing constant, allowing a $\pm$10\% variation
to adjust the position of backbending. 

The Hamiltonian is explicitly written as:
\begin{equation}
\hat{H}= \hat{H_0}-\frac{\chi}{2} \sum_ {\mu} \hat{Q}^+_{\mu}
\hat{Q}_{\mu}-G_M \hat{P}^+ \hat{P} - G_{Q} \hat{P}^+_{\mu} \hat{P}_{\mu}, 
\label{ham} 
\end{equation} 
where
\begin{equation}
\begin{array}{ll}
\hat{H_0} &= \sum_{\alpha} c^+_{\alpha} \epsilon _{\alpha} c_{\alpha} \\
\epsilon _{\alpha} &= \hbar \omega \left[ N-2 \kappa \hat{l} \cdot \hat{s}
- \kappa \mu (\hat{l}^2- < \hat{l}>^2) \right]. 
\end{array}
\end{equation}
The operators appearing in Eq. (\ref{ham}) are defined as 
\begin{equation}
\begin{array}{ll}
\hat{Q}_{\mu} &= \sum _{\alpha \beta} c^+_{\alpha} Q _{\mu \alpha \beta}
c_{\beta}\\
Q _{\mu \alpha \alpha'} &= \sqrt{\frac {4 \pi} 5} ~\delta_{N N'}
~\langle N j m | \left( {\frac r b} \right)^2 Y_{2\mu} | N' j' m' \rangle\\
\hat{P} &= \frac{1}{2} \sum _{\alpha} c^+_{\alpha} c_{\bar{\beta}}\\
\hat{P}^+ _{\mu} &= \frac{1}{2} \sum _{\alpha \beta} c^+_{\alpha} Q _{\mu
\alpha \beta} c^+_{\bar{\beta}}. 
\end{array}
\end{equation}

The quadrupole-quadrupole interaction strength will be discussed in next 
section. The monopole and quadrupole pairing interactions are given by  
\begin{equation}
\begin{array}{ll}
G_M &= (G_1 \mp G_2 \frac{N-Z}{A})\frac{1}{A}\\
G_Q &= \gamma G_M.
\end{array}
\end{equation}
The $G_M$ is inversely proportional to the particle number $A$ and it 
contains two constants $G_1$ and $G_2$. These constants are determined 
from systematics which should vary according to the size of 
the single particle space
employed in the calculation \cite{Zyman}.  

If the multi-quasiparticle basis is expressed by $ |\phi_{\kappa}>$, 
the Hamiltonian is diagonalized in the basis spanned by
$\hat{P}^I_{M K}|\phi_{\kappa}>$, where $\hat{P}^I_{M K}$ is the 
angular momentum projection operator. This lead to the
eigenvalue equation:
\begin{equation}
\sum _{\kappa ' K'} (H^I _{\kappa K \kappa 'K'}-EN^I
_{\kappa K \kappa
'K'})F^I_{ \kappa 'K'} = 0
\end{equation}
with the normalization condition:

\begin{equation}
\sum _{\kappa K \kappa 'K'} F^I_{ \kappa K}N^I _{\kappa K
\kappa 'K'}
F^I_{ \kappa 'K'} = 1, 
\label{orto}
\end{equation}
where
\begin{equation}
\begin{array}{ll}
H^I _{\kappa K \kappa 'K'} = &<\phi_{\kappa}|\hat{H}
\hat{P}^I_{ K
K'}|\phi_{\kappa'}>\\
N^I _{\kappa K \kappa 'K'} = &<\phi_{\kappa}|\hat{P}^I_{ K
K'}|\phi_{\kappa'}>. 
\label{norm}
\end{array}
\end{equation}
The normalized eigenstate is given by:
\begin{equation}
|\Psi _{IM}> = \sum _{\kappa K} F^I_{ \kappa K} \hat{P}^I_{MK}|\phi_{\kappa}>.
\label{wf}
\end{equation}

\section{Mean field deformation and the quadrupole-quadrupole force}

The logical structure of the theory is the following:

a) The Nilsson Hamiltonian is diagonalized for a given deformation.

b) The Bogolyubov transformation is performed in order of to take into
account the static monopole paring force. Three major oscillator shell are
included for protons and other three for neutrons.
It defines the Nilsson $+$ BCS quasiparticle basis. 

c) The Hamiltonian  is then  diagonalized within the shell model space
spanned by a selected set of projected multi qp-states. For even-even
nuclei we chose: 

\begin{eqnarray}
|0>, a^+ _{\nu_1} a^+ _{\nu_2}|0>, a^+ _{\pi_1} a^+ _{\pi_2}|0>, a^+
_{\nu_1} a^+ _{\nu_2}  a^+ _{\pi_1} a^+ _{\pi_2}|0>.
\end{eqnarray} 

The effect of rotation is described by the projection operator and the
whole dependence of the wave functions on spin is contained in the
eigenvectors  since the quasiparticle basis is spin independent. It is
this
feature which makes not only the treatment simple and stable but also the
interpretation of the result easy and intuitive.  

The quadrupole-quadrupole interaction strength is obtained identifying 
the Hartree potential resulting from Hamiltonian (\ref{ham}) with
the Nilsson stretched potential 
\begin{equation}
\begin{array}{ll}
H_{Nilsson}^\alpha = H_0^\alpha - {\frac 2 3} \epsilon_\alpha \hbar
\omega_\alpha \hat Q_0 ,
\hspace{1.5cm}\alpha = p,n , ~~~\\
\omega_\alpha = \omega_0 \{ 1 \pm {\frac {N-Z} A} \}^{\frac 1 2}, 
\hspace{1.5cm}
\hbar \omega_0 = 41.47 ~A^{- {\frac 1 3}} ~MeV , 
\label{Nilsson}
\end{array}
\end{equation}

\noindent
with the + ($-$) sign holding for $\alpha$ = neutron (proton).
 
The HFB single-particle Hamiltonian which results from Eq. (\ref{ham}) is:

\begin{equation}
\hat{H} = \hat{H_0} - \chi <\hat{Q}_{0}> \hat{Q}_{0}-G <\hat{P}>(
\hat{P}+\hat{P}^+). 
\label{hfb}
\end{equation}
Equating the first two terms in (\ref{hfb}) with the Nilsson Hamiltonian
(\ref{Nilsson}) we obtain the self-consistent condition

\begin{equation}
\begin{array}{ll}
\chi _{nn}<\hat{Q_0}>_n + \chi _{np}<\hat{Q_0}>_p = &\frac{2}{3}\hbar\omega
_n\epsilon,\\
\chi _{pp}<\hat{Q_0}>_p + \chi _{pn}<\hat{Q_0}>_n = &\frac{2}{3}\hbar\omega
_p\epsilon
\end{array}
\end{equation}

\noindent
where the same deformation $\epsilon$ was used for protons and neutrons.

Assuming a pure isoscalar coupling \cite{Har95}, i.e. 
\begin{eqnarray}
\chi _{np} = \chi _{np} = \chi_{pp} {\frac {\omega_n} {\omega_p}} =
\chi_{nn} {\frac {\omega_p} {\omega_n}}
\end{eqnarray}

\noindent
the above equations determine the Q-Q
coupling constants $\chi$ as a function of $\epsilon$:

\begin{equation}
\chi_{\alpha \alpha'} = {\frac
{ {\frac{2}{3}} \epsilon \hbar\omega_\alpha \hbar\omega_{\alpha'}   }
{ \hbar\omega_n <\hat{Q_0}>_n + \hbar\omega_p<\hat{Q_0}>_p }
} 		\label{chi}
\end{equation}
The expectation values $<\hat{Q_0}>_n$ and 
$<\hat{Q_0}>_p$ are evaluated adding
the quadrupole moments of all active deformed Nilsson orbitals weighted by
their occupation numbers. They depend, of course, on the assumed
mean field deformation $\epsilon$ and vanish for spherical nuclei, leaving
$\chi$ undetermined in that case. 

\section{Variation of the parameters}

While in the previous sections it was argued that all the parameters in the
Hamiltonian (1) are fixed, either by self-consistency
(quadrupole-quadrupole) or by systematics (mean field deformation,
pairing), it is well known that the truncation of the Hilbert space affect
those parameters. As an example, when in RPA calculations the ``physical''
quadrupole operator is replaced by its ``algebraic'' counterpart , which do
not mix different major shells, the self-consistent $\chi$ is usually
enlarged by a factor of two to compensate for this truncation in order to get
the right spectra \cite{Bes69}. In this section we will study the influence
on the spectra when 
the residual interaction strengths depart from their
self-consistent values.

With the value of the quadrupole-quadrupole strength  $\chi$
self-consistently determined from Eq. (\ref{chi}), we introduce two
adimensional parameters $x$ and $g$ in the Hamiltonian, which now looks like

\begin{equation}
\hat{H}= \hat{H_0}- x \frac{\chi}{2} \sum_ {\mu} \hat{Q}^+_{\mu}
\hat{Q}_{\mu}- g G_M \hat{P}^+ \hat{P} - G_{Q} \hat{P}^+_{\mu}
\hat{P}_{\mu}.
\label{hamxg}
\end{equation} 
When $x=g=1$, we recover the original Hamiltonian (\ref{ham}). 

In what follows we will present calculations for $^{160}Dy$. The BCS
equations were solved for Nilsson single particle energies including the
$N= 3,4,5$ major shells for protons and the $N= 4,5,6$ for neutrons.
The reported deformation $\epsilon = 0.25$ \cite{ADT95} is used, but some
results calculated with $\epsilon = 0.29$ are also presented.
Proton and neutrons quasiparticle pairs with energies lower than 1.5 MeV,
as well as states with two proton and two neutron quasiparticles with
energies lower than 3.0 MeV were 
included in conjunction with the BCS vacuum to
build up the truncated Hilbert space.

In Fig. 1, the band-head energies are plotted against $x$, when no
residual pairing and quadrupole-pairing interactions are present ($g =
0$ in (\ref{hamxg}), $\gamma = 0$ in (4)). The notation used to classify
the states is $J_i$, where $J$ is the bandhead angular momentum and $i$
enumerates the states with the same $J$ in order of increasing energy. On
the right corner of Fig. 1, the
experimental energies are given. 
It is clearly seen that there is only a narrow
region around $x \approx 1.0$ where all the states are compressed, in good
agreement with the experimental data. Beyond this region the energies of
two excited $0^+$ and two $4^+$ states grow more than four times,
providing the image of a valley whose bottom reproduce the experimental
energies. The first excited $0^+$ state ($0_2$) is nearly insensitive to
changes in
$x$. The first $1^+$ state exhibit a change in slope when $x \approx1.3$.
In $^{160}Gd$ the bandheads with $J=1$ and $J=4$ are nearly degenerated,
and there is some consensus that they are built mainly from the same qp
pair, which has $k_1 + k_2 = 4, ~|k_1-k_2| = 1$. Only in a small region
around $x \approx 1.0$ these two states ($1_1$ and $4_2$) remain close to
each other. In this region their dominant components are the two
quasi-neutron states [5 f7/2 -3/2]n and [5 h9/2 5/2]n.

In Table 1 we list the main components of each bandhead state for four
values of $x$. It is remarkable that only for $x=1$ all of them can be
described as two quasiparticle bands, while in all the other cases some
states have important four quasiparticle components.
Notice also that for small $x$ values there is a crossing between $4_1$
and $4_2$ and their interchange their components. 

The first $2^+$ state is usually assigned as 
the bandhead of the gamma band, which is 
interpreted as a rotor band built over a one-phonon quadrupole vibration.
In Fig. 1, it can be seen that the
first $2^+$ state has a complicated behaviour and its
energy at $x=1$ is nearly twice the
experimental value $E_{\gamma}{(2^+)_{exp}}=0.974$ MeV. This results is not
privative of $^{160}Dy$. In many other deformed nuclei the PSM predicts
a first $2^+$ band too high in energy \cite{Har95}. 
For $x > 1.4$ the energy of this state is decreasing, and around $x
\approx 2.4$ it would become the ground state.
A microscopic analysis of the gamma band is presented in the
following section.

In order to check the validity of the above mentioned results we have
performed a similar calculation with a different deformation. We chose
$\epsilon=0.29$ rather than the reported $\epsilon=0.25$  which provides a
better fit of the rotational ground state band. We will discuss the
rationale of selecting a larger deformation for this particular nucleus
in section 6. By the moment the relevant point is to confirm that only
around the self-consistent values $x=1$ the bandheads energies reproduce  
their experimental values. They are plotted in Fig. 2 against $x$, again
for $g=0, \gamma = 0$. We found that 
curves are similar to those shown in Fig. 1,
exhibiting a deep valley. The gamma band is still to high, but now the first
excited $4^+$ state is going down in energy faster than the $2^+$ for
$x > 1.2$. The origin of this difference can be traced back to the
quasiparticle content of both bands. A change in deformation implies a
different active valence particle region, with other two quasiparticle pairs
commanding the behaviour of the low energy apart of the spectra.

In Fig. 3 we present a similar analysis including both monopole and
quadrupole pairing ($g=1, \gamma=0.18$). It shows that the inclusion
of the remaining interactions do not introduce significant changes. It
means that the gross structure of the bandheads is due to the
mean field and quadrupole-quadrupole interaction and that the residual 
pairing and quadrupole-pairing interactions just induce minor
rearrangements in the bandhead positions. Pairing strengths significantly
influence the level spacing within a band. 

If isovector couplings are allowed, violating relations (13), a similar
trend is found. It is shown in Fig. 4, where we kept $\chi_{pp}$ and
$\chi_{nn}$ at their self-consistent values but renormalized
$\chi_{pn}=x_3\chi^0_{pn}$. The energy variations are smaller than in the
previous figures because we are only changing one component of the
quadrupole-quadrupole force.

Fig. 5 shows the bandhead energies as function of the pairing strength
parameter $g$ for $x=1, \gamma =
0.18$. All curves are smooth and grow monotonically with $g$. It provides
additional evidence about the dominant role of the residual
quadrupole-quadrupole force in shaping the spectra, while the residual
pairing interaction has a very modest effect in the low energy spectra.

\section{The gamma band}

The major components $F^{I=2}_{ \kappa K}$ of the gamma bandhead are
plotted in Figs. 6 and 7 against their two-quasiparticle energies for $x =
0.7$ and $1.4$ respectively. Near each line the pair angular momentum
projection in the intrinsic system $K = k_1 \pm k_2$ is given, as well as
the two quasiparticle pairs they are built of. 
Notice that the components $F^{I=2}_{ \kappa K}$ of the  wave functions
(\ref{wf}) are not orthonormal in the usual sense, but satisfy the
orthonormality relationships (\ref{orto}), where the overlap matrix
(\ref{norm}) can have very small components. For this reason the
wave function components can be larger than one, and the relevant
information if Figs. 6 and 7 lies in the relative magnitudes and phases.

For $x=0.7$ a single quasiparticle pair ([6 i13/2 1/2]n-[6 i13/2 -3/2]n)
 with $K=2$ with
qp-energy slightly smaller than 1 MeV dominates the $2^+$ state which is
the head of the gamma band, with small mixing with two $K=0$, two $K=1$
and another $K=2$ quasiparticle pairs. In contrast, for $x=1.4$ we have at
least nine $K=2$ qp pairs which contributes, mainly in a coherent way, to
the gamma bandhead wave function.

Qualitatively we would expect the gamma band to have a collective
character, as that found for $x=1.4$. On the other side, the energy of the
gamma bandhead is always to high, as pointed out in the previous section.
Enlarging the Hilbert space do not change this result. It seems that the
wave function has enough collectivity but the Hamiltonian is not rich
enough to fit its energy correctly.

In Fig. 8 the variation of the energy of the gamma bandhead is presented as
function of the mean field deformation for $0.16 \le \epsilon \le 0.32$.
All the other Hamiltonian parameters were fixed self-consistently. It can be
seen that the energy of this $J=2$ state decrease in a smooth monotonic
way, but the change is too small to adjust the experimental energy.

The above discussion about the gamma bandhead lead us to conclude that
there is a clear need to include additional degrees of freedom in
Hamiltonian (\ref{ham}) 
in order to allow the PSM to describe such fine details of
the spectra. 
Changes in the strength of the quadrupole-quadrupole or
pairing forces would have only strongly negative effects on the general
description of the spectra. 
Through many studies in the past, the presence of these forces 
in a Hamiltonian has
been found to be very important, if not sufficient, to explain a
large body of data. They simulate the essence of the most important 
correlations in nuclei, so that even the realistic force has to
contain at least these components implicitly in order for it to
work successfully \cite{Zuker96}. Of course,  
the simple pairing plus quadrupole forces cannot describe everything
that is taken place in many-nucleon systems and thus have to be
supplemented with other types of interaction whenever neccessary. 
Spin dependent two body interactions, which
have important contributions in other microscopic theories \cite{Sol86}, 
could be possible candidates to enrich the PSM Hamiltonian.

\section{The $^{160}Dy$ yrast band}

The PSM does an impressive job in describing the spectra
and electromagnetic transitions in heavy deformed \cite{Har95} 
and superdeformed nuclei
\cite{Sun97}. However, for the  $Dy$ isotopes some discrepancies has been
found in previous works\cite{Sun94}.

We present in Fig. 9 the alignment diagram of the yrast band in $^{160}Dy$.
It plots energy differences $\omega = (E(J) - E(J-2))/2$ against the
angular momentum J, for two different deformations $\epsilon = 0.25$ and
$0.29$. The experimental data are shown as a thick line. It is apparent
that the larger deformation is able to reproduce the data.

The backbending plot showed in Fig. 10 reinforces this conclusion. The graph
presents twice the moment of inertia $\Theta$ \cite{Har95} as function of
$\omega^2$ for the same two deformations. The backbending is exaggerated for
$\epsilon = 0.25$ and well reproduced by the smoother curve with $\epsilon
=
0.29$. Therefore, an 15
\% increase in the deformation allows the PSM to match the experimental
data. It deserves a closer and more systematic analysis over many isotopes,
including B(E2) transitions, which will be presented elsewhere \cite{Vel98}. 

\section{Summary}

Properties of the spectra of $^{160}Dy$ in the Projected Shell Model were
studied allowing an artificial variation of the parameter strengths from
their self-consistent values (in the case of the quadrupole-quadrupole
interaction) or their values taken from systematics (Nilsson mean field
deformation and pairing). 

The dominant role of the quadrupole-quadrupole force in deformed nuclei was
once again confirmed. It was clearly exhibited that any departure of its
strength from the value obtained self-consistently from the mean field
deformation completely destroy the agreement with experimental data,
because many excited bands increase drastically their energies. 
It was concluded that, even when a strongly truncated 
quasiparticle space is in use,
the quadrupole-quadrupole force in the PSM is completely determined by
self-consistency. It is worth mentioning that the same self-consistent
approach has been also fruitful to remove spurious states in the RPA
treatment of deformed nuclei \cite{LoI96}.

The residual pairing interaction between quasiparticles has only a smooth
effect over the spectra. It provides a better description of excited bands
but is just a second order correction.

The mean field deformation $\epsilon$ is in principle fixed by systematics
\cite{ADT95}. However, we found that, for $^{160}Dy$, using a 15 \% larger
deformation provides a notably improved description of backbending. A
detailed study of the spectra an electromagnetic transitions in $Dy$
isotopes using a larger deformation will be presented in another article.

We also found the limitations of the PSM in the description of the gamma
band. It was demonstrated that Hamiltonian (1) would need to be improved to
be able to correctly describe the energetics to these states. Inclusion of
a spin-dependent interaction could work in this direction.

\section{Aknowledgements}

This work was supported in part by Conacyt (Mexico) and the National
Science Foundation.

\newpage

\begin{table}
\begin{center}
\caption{ 
Dominant spherical quasiparticle component of each bandhead state
for 4 values of $x$. The $x$ values, $J_i$ banhead classification,
energies of the 2-qp pairs, $K$ angular momentum projection and the
spherical labels of the qp pairs are listed.
}
\begin{tabular} {|c|c|c|c|c|}
\hline
$x$ & Bandhead & Pair Energy & Projection & Spherical labels [N L J
M]$T_3$ \\ \hline
0.0 & $0_2$ & 3.6525 & 0 & 4qp \\
    & $0_3$ & 2.2842 & 0 & [5 h11/2 5/2]p[5 h11/2 5/2]p \\
    & $0_4$ & 1.4283 & 0 & [5 f7/2 -3/2]n [5 f7/2 -3/2]n\\
    & $1_1$ & 3.8841 & -1 & 4qp \\
    & $2_2$ & 2.4481 & 2 & [6 i13/2 1/2]n [6 i13/2 -3/2]n \\
    & $4_2$ & 1.7536 & -4 & [6 i 13/2 3/2]n [6 i13/2 5/2]n \\
    & $4_3$ & 1.4204 & -4 & [5 f7/2 -3/2]n [5 h9/2 5/2]n \\
\hline
0.7 & $0_2$ & 3.6525 & 0 & 4qp \\
    & $0_3$ & 1.8315 & 0 & [6 i13/2 -3/2]n [6 i13/2 -3/2]n \\ 
    & $0_4$ & 1.4283 & 0 & [5 f7/2 -3/2]n [5 f7/2 -3/2]n \\
    & $1_1$ & 3.8841 & -1 & 4qp \\
    & $2_2$ & 2.4481 & 2 & [6 i13/2 1/2]n [6 i13/2 -3/2]n \\
    & $4_2$ & 1.4204 & -4 & [5 f7/2 -3/2]n [5 h9/2 5/2]n \\
    & $4_3$ & 1.7536 & -4 & [6 i13/2 -3/2]n [6 i13/2 5/2]n \\
\hline
1.0 & $0_2$ & 1.8211 & 0 & [5 h11/2 -7/2]p [5 h11/2 -7/2]p\\
    & $0_3$ & 1.4126 & 0 & [5 h9/2 5/2]n [5 h9/2 5/2]n\\
    & $0_4$ & 1.6758 & 0 & [6 i13/2 5/2]n [6 i13/2 5/2]n\\
    & $1_1$ & 1.4204 & 1 & [5 f7/2 -3/2]n [5 h9/2 5/2]n \\
    & $2_2$ & 2.0985 & -2 & [4 d5/2 -3/2]p [4 d3/2 1/2]p \\
    & $4_2$ & 1.4204 & -4 & [5 f7/2 -3/2]n [5 h9/2 5/2]n \\
    & $4_3$ & 1.7536 & -4 & [6 i13/2 -3/2]n [6 i13/2 5/2]n  \\
\hline
1.4 & $0_2$ & 3.4968 & 0 & 4qp \\
    & $0_3$ & 1.8315 & 0 & [6 i13/2 -3/2]n [6 i13/2 -3/2]n\\
    & $0_4$ & 3.6525 & 0 & 4qp \\
    & $1_1$ & 3.7284 & -1 & 4qp \\
    & $2_2$ & 4.0111 & 2 & [6 i13/2 5/2]n [6 d5/2 1/2]n \\
    & $4_2$ & 1.4204 & -4 & [5 f7/2 -3/2]n [5 h9/2 5/2]n \\
    & $4_3$ & 3.5747 & -4 & 4qp \\
\hline
\end{tabular}
\end{center}
\end{table}

\newpage

\centerline{\bf Figure Captions}

\bigskip

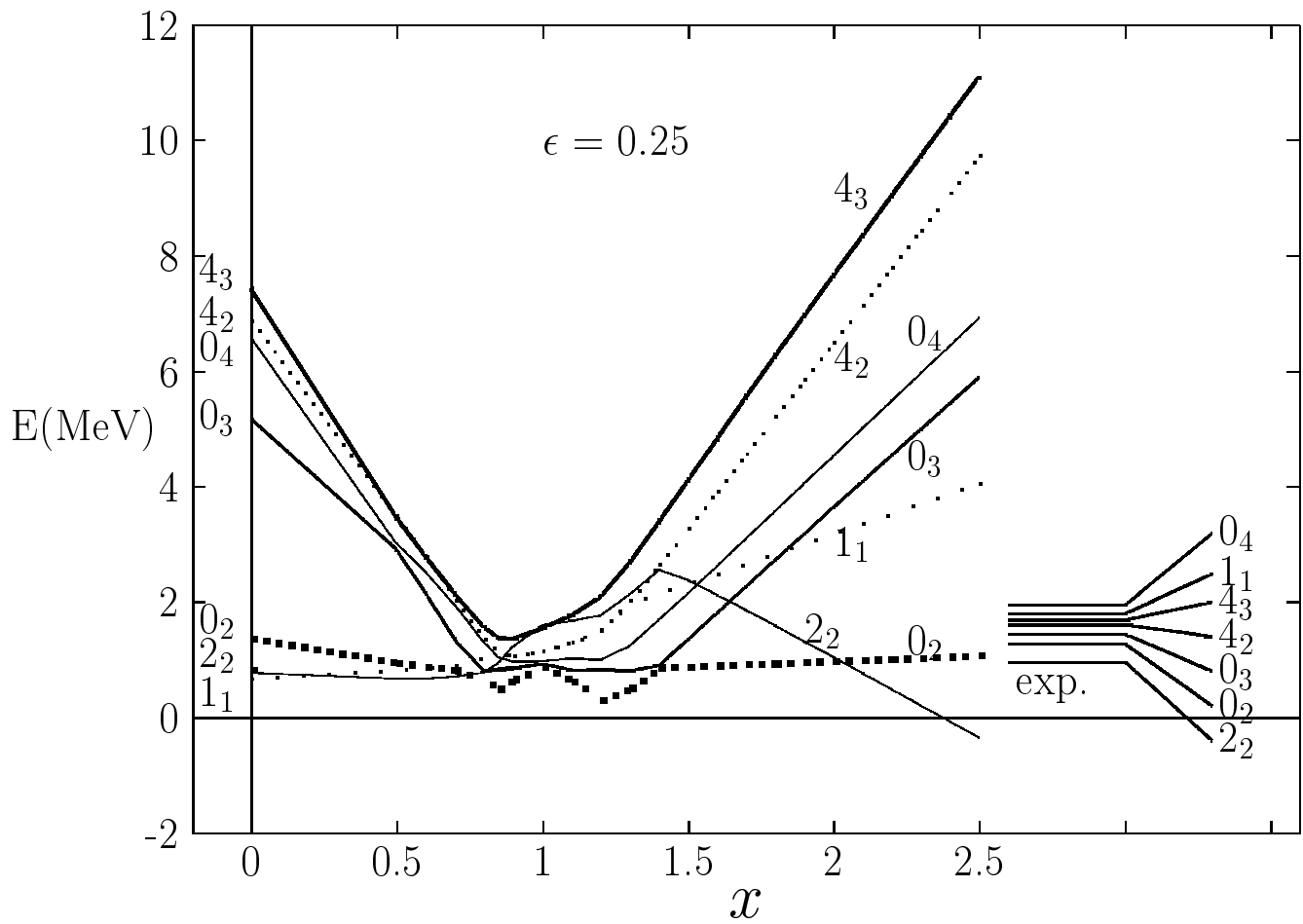
Figure 1: Bandheads energies vs. $x$ for $\epsilon = 0.25, g=0, \gamma
=0$. The notation $J_i$ is used to indicate the angular momentum $J$ and the
order $i$, in increasing energy, of each state. On the right hand side the
experimental energies are drawn.
 
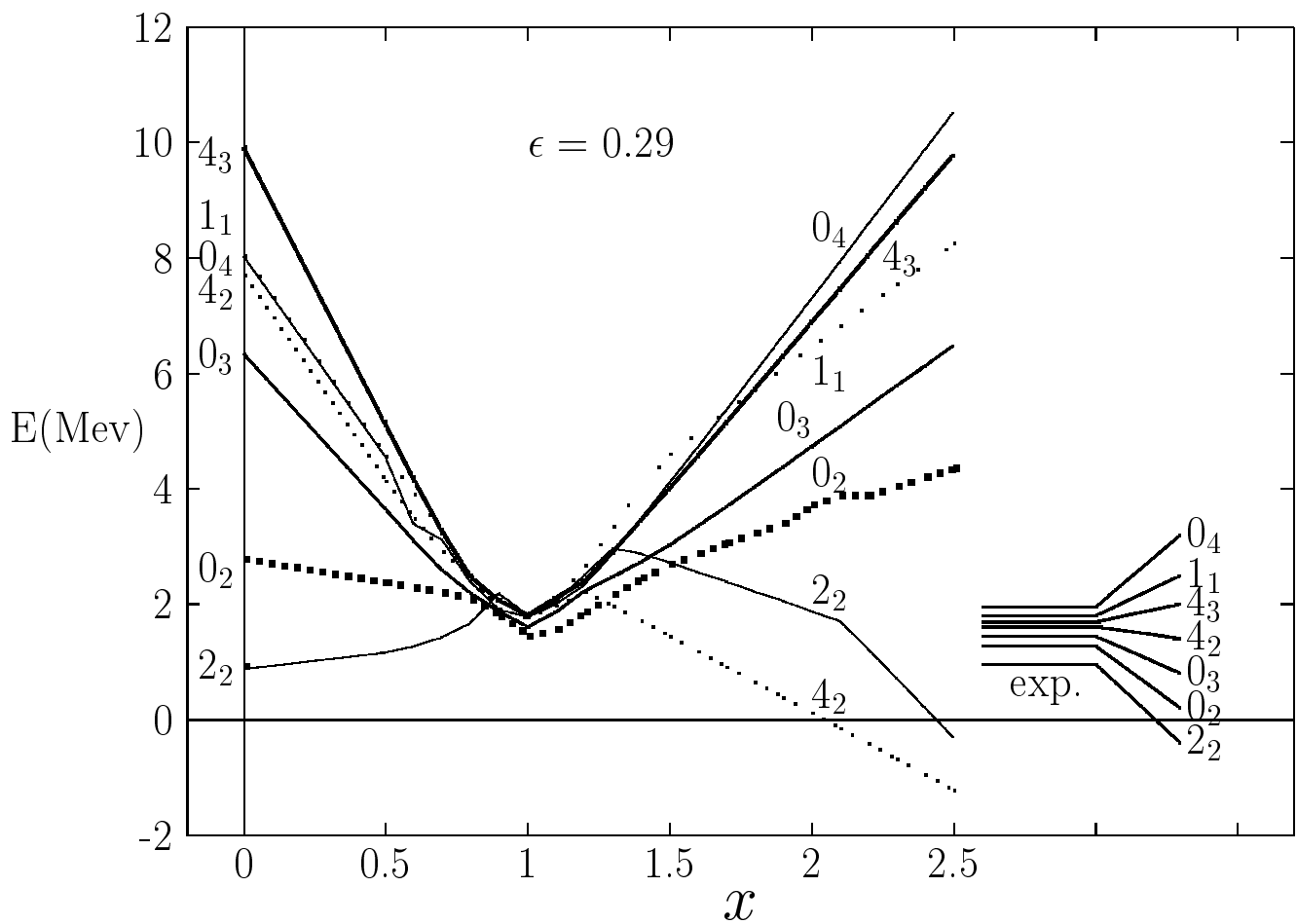
Figure 2: Bandheads energies vs. $x$ for $\epsilon = 0.29, g=0, \gamma
=0$.

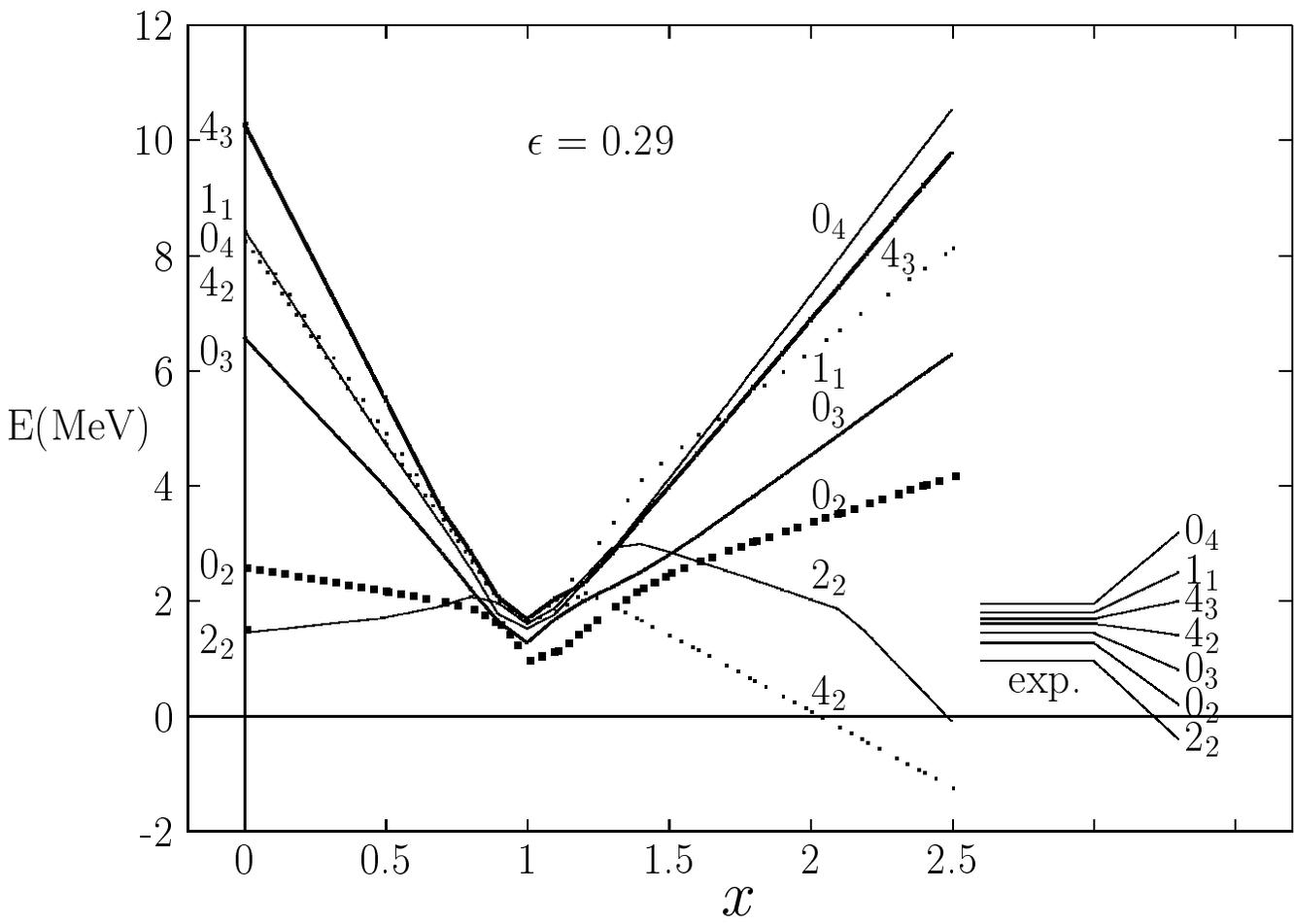
Figure 3: Bandheads energies vs. $x$ for $\epsilon = 0.29, g=1.0, \gamma
=0.18$ .

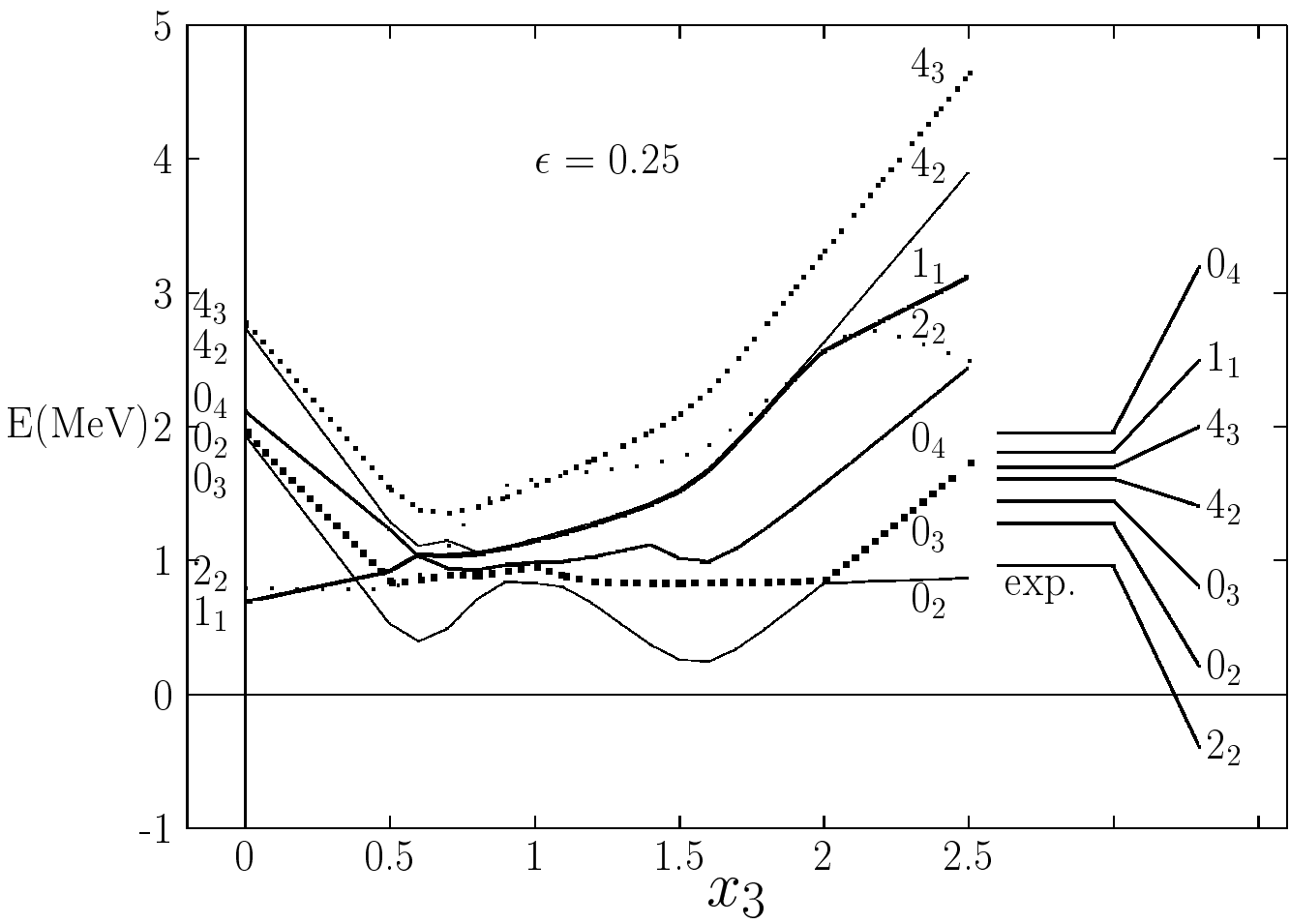
Figure 4: Bandheads energies vs. $x_3$ for $\epsilon = 0.25, g=1.0, \gamma
=0.18$ .

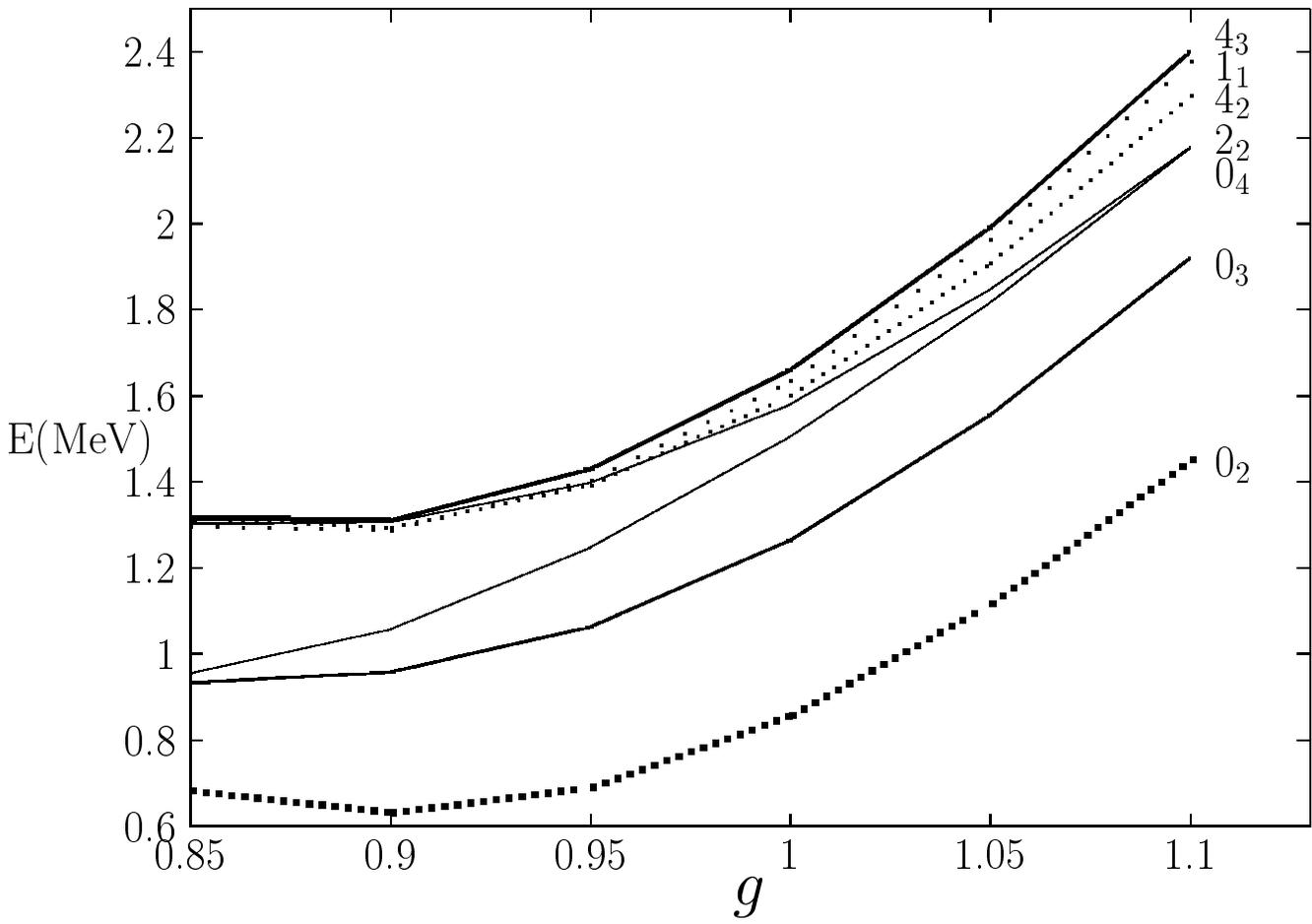
Figure 5: Bandheads energies vs. $g$ for $\epsilon = 0.29, x=1.0, \gamma
=0.18$ . 

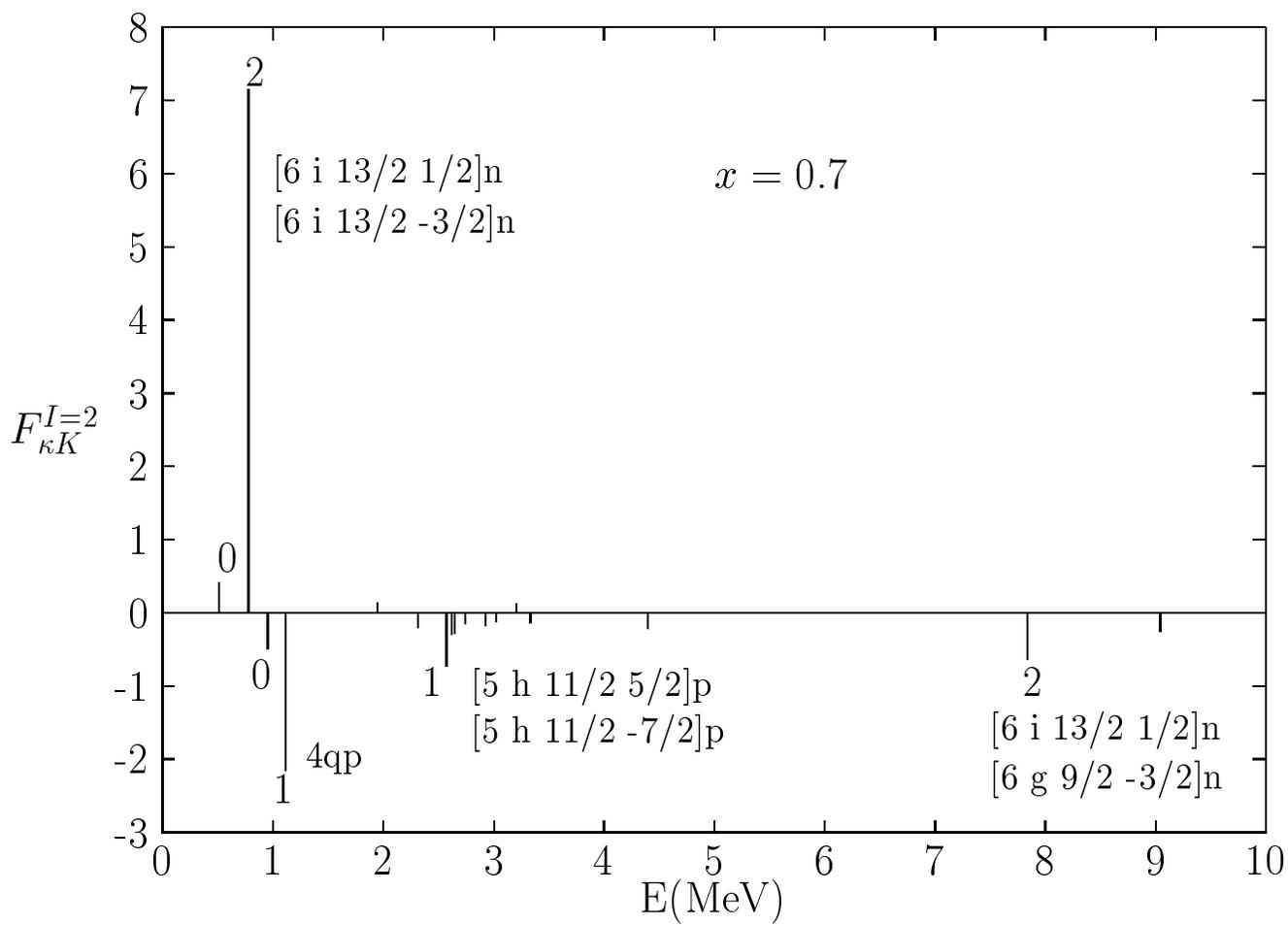
Figure 6: Gamma bandhead main components for $x=0.7$ as a function of the
2qp-energy. The angular momentum projection $K = k_1 \pm k_2$ is given
near each line, together with the main 2 qp components. 4 qp means that
that a four quasiparticle pair dominates the state.

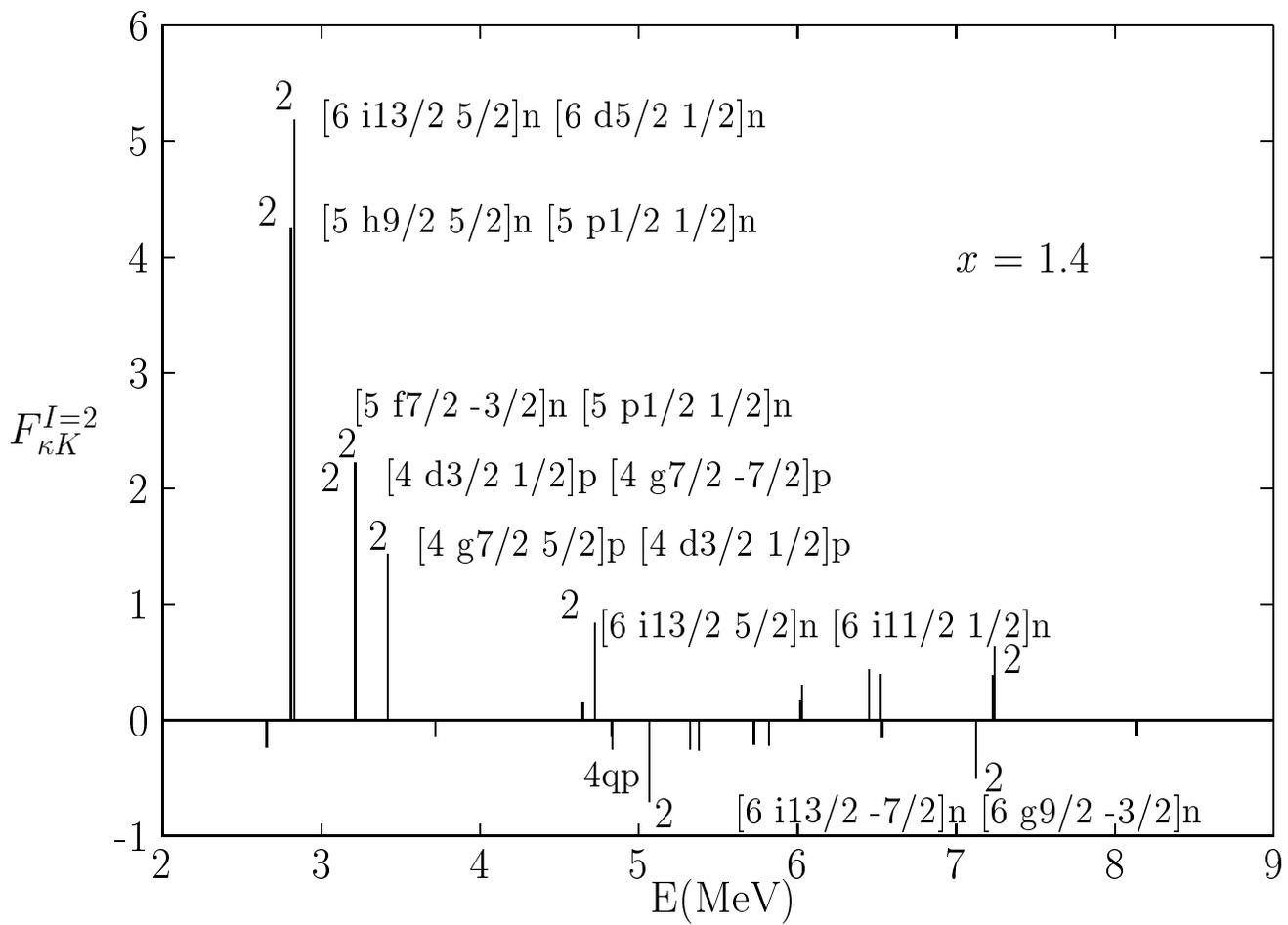
Figure 7: Gamma bandhead main components for $x=1.4$ as a function of the
2qp-energy.

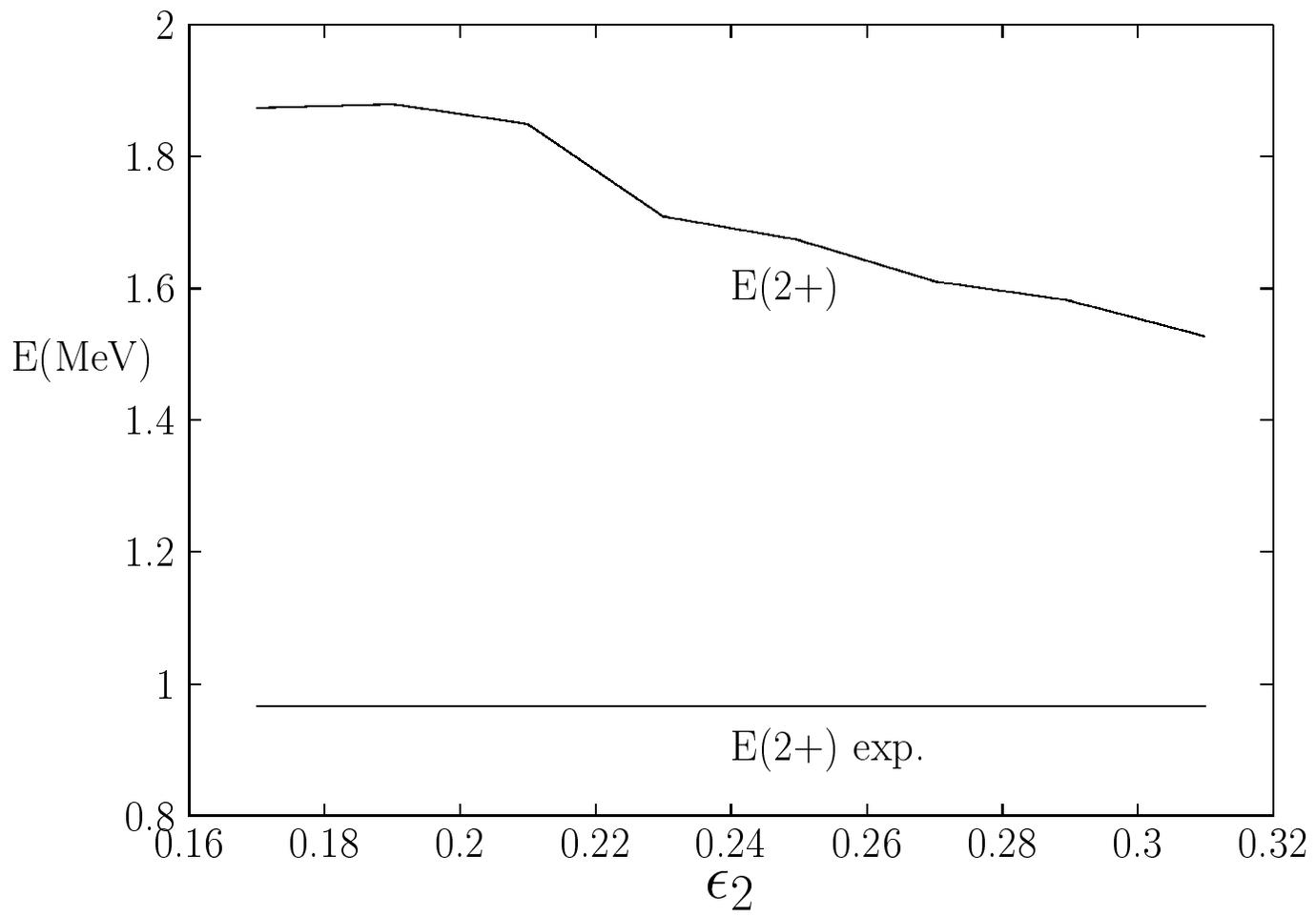
Figure 8: Gamma bandhead energy vs. $\epsilon$.

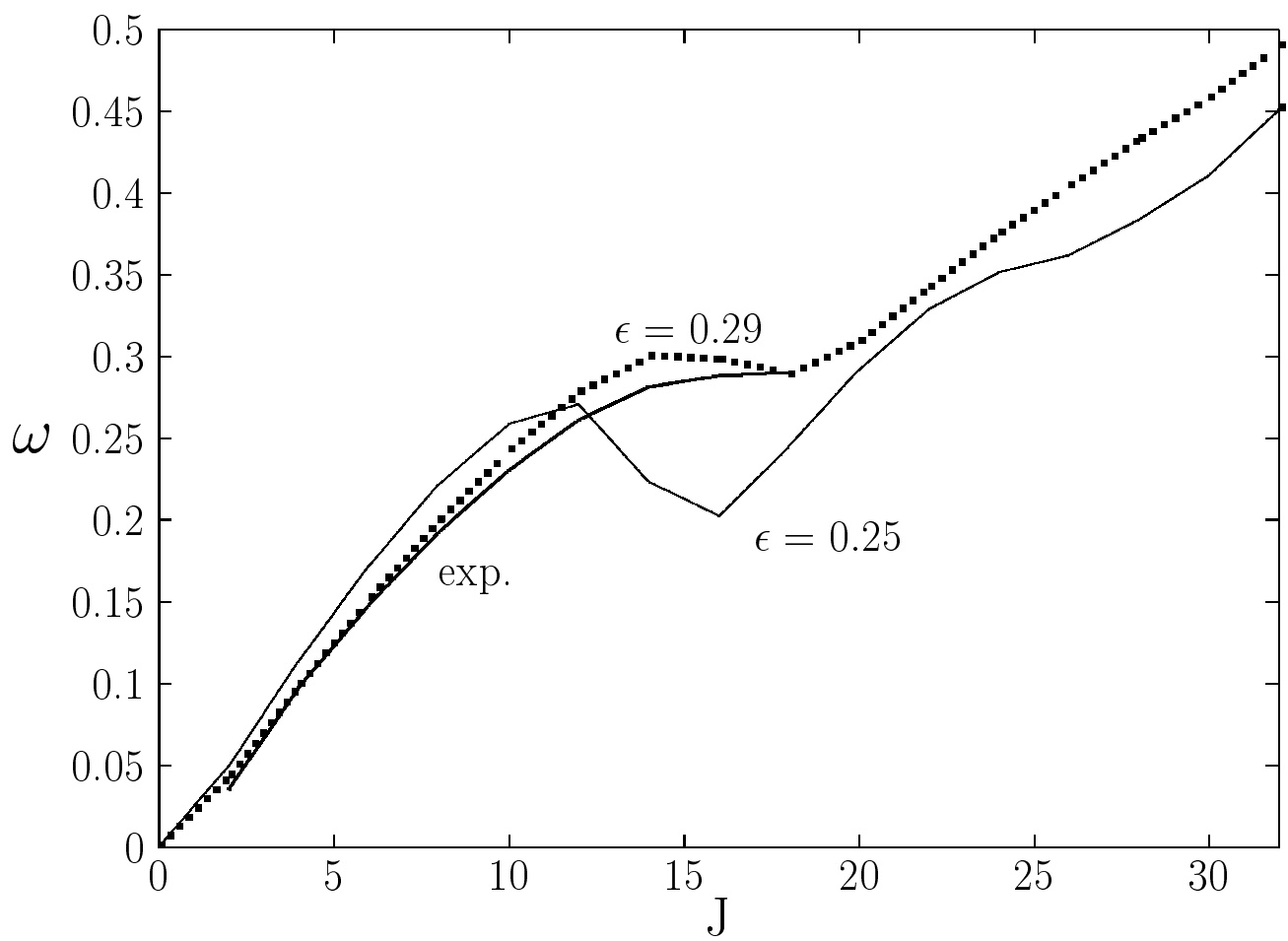
Figure 9: Angular frequencies versus the angular momentum in $^{160}Dy$. 

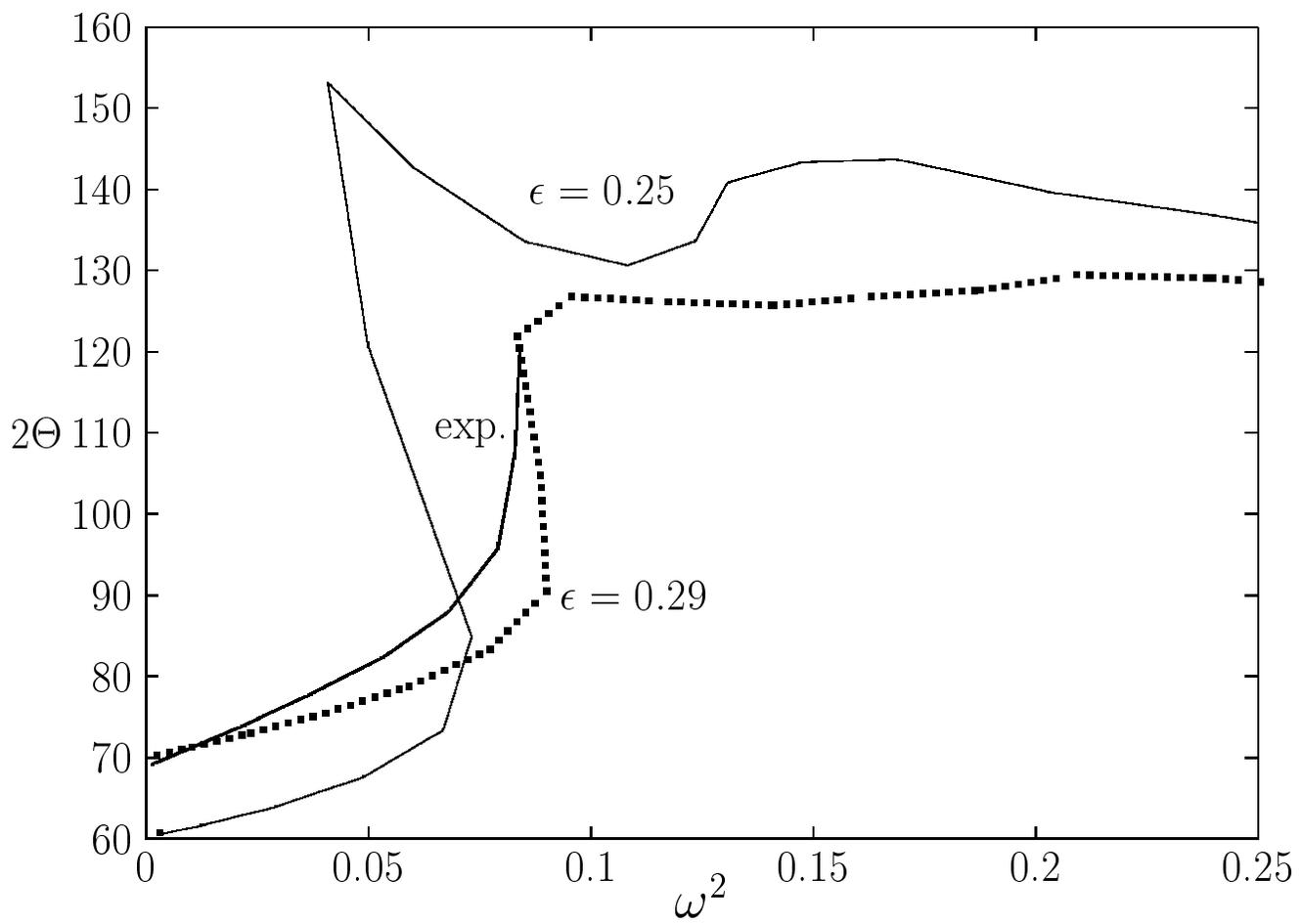
Figure 10: The moment of inertia as function of the squared angular
frequency for $^{160}Dy$. 

\end{document}